\documentclass{sig-alternate}

\usepackage{amssymb}
\usepackage{amsmath}
\usepackage{algorithm}
\usepackage{algorithmic}
\usepackage{graphicx}
\usepackage{epstopdf}
\usepackage{color}
\usepackage{balance}
\usepackage{wrapfig}
\usepackage{caption}
\usepackage{subcaption}

\begin{document}

\conferenceinfo{Technical Report}{ArXiv}

\title{MonoStream: A Minimal-Hardware High Accuracy Device-free WLAN Localization System}

\numberofauthors{2}

\author{
\alignauthor
Ibrahim Sabek\\
       \affaddr{Comp. and Sys. Eng. Dept.}\\
       \affaddr{Alexandria University, Egypt}\\
       \email{ibrahim.sabek@alexu.edu.eg}
\and
\alignauthor
Moustafa Youssef\\
       \affaddr{Wireless Research Center}\\
       \affaddr{E-JUST, Egypt}\\
       \email{moustafa.youssef@ejust.edu.eg}
}

\maketitle

\begin{abstract}
Device-free (DF) localization is an emerging technology that allows the detection and tracking of entities that do not carry any devices not participate actively in the localization process. Typically, DF systems require a large number of transmitters and receivers to achieve acceptable accuracy, which is not available in many scenarios such as homes and small businesses. 

In this paper, we introduce \textit{MonoStream} as an accurate single-stream DF localization system that leverages the rich Channel State Information (CSI) as well as MIMO information from the physical layer to provide accurate DF localization with only one stream. To boost its accuracy and attain low computational requirements, \emph{MonoStream} models the DF localization problem as an object recognition problem and uses a novel set of CSI-context features and techniques with proven accuracy and efficiency.
Experimental evaluation in two typical testbeds, with a side-by-side comparison with the state-of-the-art, shows that \textit{MonoStream} can achieve an accuracy of $0.95 m$ with at least $26$\% enhancement in median distance error using a single stream only. This enhancement in accuracy comes with an efficient execution of less than $23ms$ per location update on a typical laptop. This highlights the potential of \emph{MonoStream}  usage for real-time DF tracking applications.
\end{abstract}




\newpage
\section{Introduction}
\label{sec:intro}

\begin{figure}[!t]
\centering
\includegraphics[width=3in]{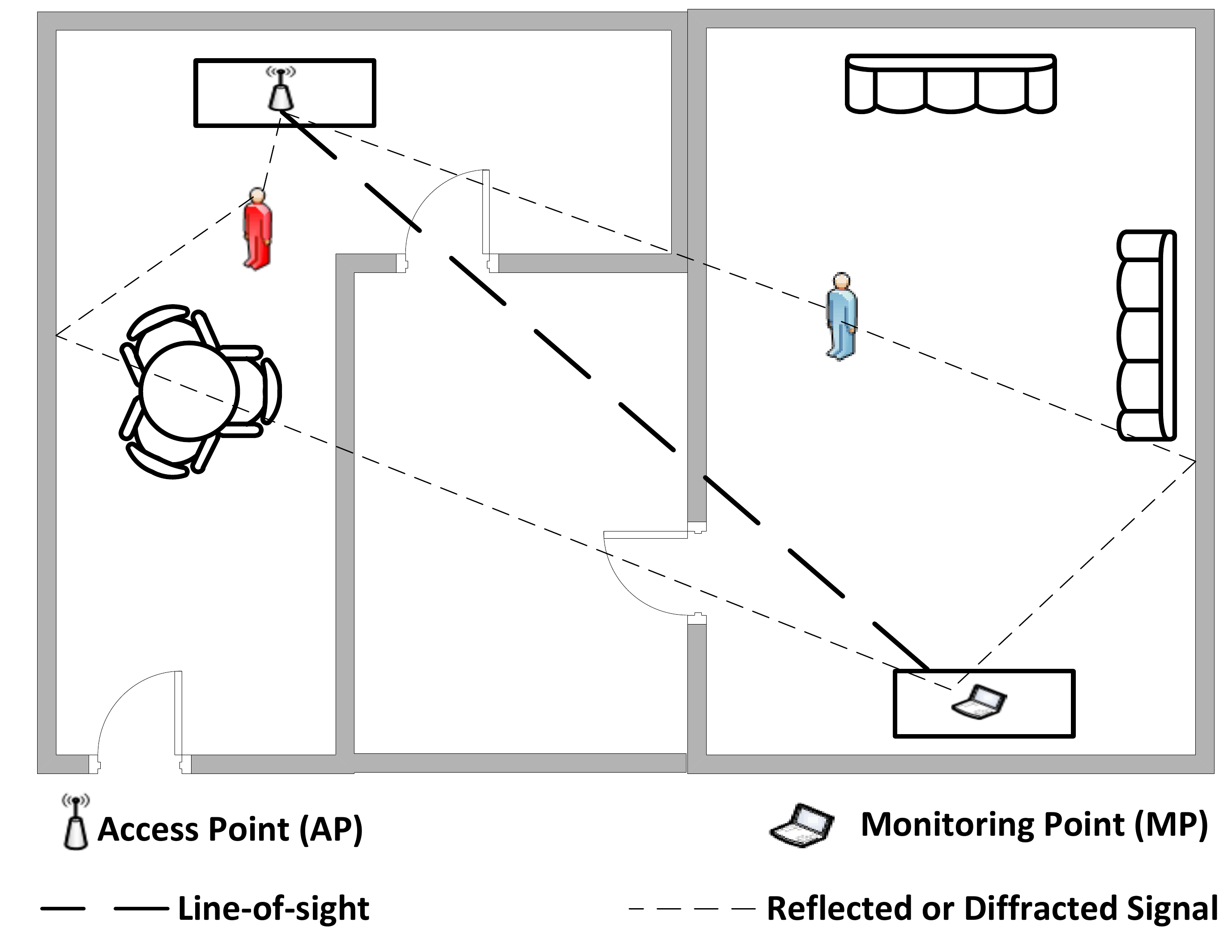}
\caption{Typical architecture of a DF WLAN localization system.}
\label{fig:dfp_arch}
\end{figure}


Many localization systems have been proposed over the years including the GPS system \cite{Enge:GPS}, RF-based systems \cite{Bahl:radar,Youssef:Horus}, and infrared-based systems \cite{Want:Badge}. All these systems require that the tracked entity carries a device.
On the other hand, device-free (DF) passive localization \cite{Youssef:DFPchallenges} is based on using typical wireless networks to detect and track entities that do not carry any devices nor participate actively in the localization process. It depends on the fact that the RF signal strength is affected by human motion. DF localization can be used in many applications including smart homes, intrusion detection, medical care, and traffic estimation. A typical DF system (Figure~\ref{fig:dfp_arch}) consists of signal transmitters (such as standard access points (APs)), monitoring points (MPs) (such as standard laptops or APs themselves), and an application server for processing the received signal strength from the MPs to detect and localize events.

Current approaches for DF localization include radar-based systems, e.g. \cite{Yang:UWB,Lin:Doppler,Haimovich:MIMO}, computer vision based systems, e.g. \cite{Moeslund:Camera,Krumm:Camera} and Radio Tomographic Imaging (RTI), e.g. \cite{Patwari:RTI}. These systems, however, need special hardware and high installation cost. On the contrary, a number of DF localization systems have been proposed that operate in standard WiFi networks, e.g. Nuzzer \cite{Nuzzer_TMC} and Rasid \cite{Kosba:RASID}, without requiring any additional equipment. Therefore, these systems provide a value added-service on top of the wireless infrastructure, just based on the reported signal strength from the MAC layer. \textbf{\emph{While these systems provide lower cost localization services, they still require a large number of streams (a data stream is the data received from one AP at one MP), which limits their applicability and accuracy in a large class of scenarios, such as in homes and small bussinesses, where usually a small number, typically one AP is installed.}}

In this paper, we introduce \textit{MonoStream} as a high-accuracy limited-hardware device-free WLAN localization system.  \textit{MonoStream} is designed to work with a low number of streams, typically one (i.e. one AP and one MP). To compensate for the loss of information due to reducing the number of streams, \textit{MonoStream} leverages the detailed physical layer information of WiFi networks. In particular, the IEEE 802.11n standard uses the OFDM modulation, where a wide channel is divided into several orthogonal sub-carriers each arriving at the location of the receiver with distinct values of phase and magnitude (denoted as Channel State Information
(CSI)). This provides rich information to detect the effect of human motion on the magnitude of each subcarrier, as compared to a single signal strength value that has been used with the current approaches. In addition, the IEEE 802.11n devices use the MIMO technology, which further provides more information about each antenna pair from the transmitter to the receiver.

To further address the noisy wireless channel and the rich CSI information, \textit{MonoStream} models the DF localization problem as an object recognition problem, where it treats the CSI profile at different locations as images and extracts novel features that can capture \emph{small variations} in the effect of the human standing at different locations in the area of interest on the CSI vectors. To reduce the computational cost of the proposed techniques on the application server, \textit{MonoStream} employs a joint boosting technique \cite{Torralba:Sharing04} to scale up for a large number of features and locations.
Experimental evaluation, in  two typical testbeds using a single access point and a single laptop, shows that \textit{MonoStream} can achieve a localization accuracy of less than 0.95m. This corresponds to at least 26\% enhancement in median distance error over the state-of-the-art DF localization systems using the same environment. This comes with a real-time location update rate that requires less than $23ms$ per location estimate.

The rest of the paper is organized as follows: Section \ref{sec:background} introduces a brief background about the physical layer information and its properties that can be used to identify the human location.  Following that, Section \ref{sec:mono_stream} presents the \emph{MonoStream} system details. Section \ref{sec:eval} describes the performance evaluation of the proposed system, comparing it to the state-of-the-art DF localization systems. Then, 
we present the related work in section \ref{sec:related}, and discuss the  other points related to the system in section \ref{sec:discussion}. Finally, we conclude the paper and give directions for future work in section \ref{sec:conclusion}.


\section{Background and CSI Characterization}
 \label{sec:background}
In this section, we briefly present the OFDM modulation and MIMO technologies as well as the basic information we rely on to build the core system blocks.

\subsection{Channel State Information (CSI) and MIMO Technology}
  Many IEEE 802.11 standards (e.g. a/g/n) use OFDM modulation that transmits signals over several orthogonal frequencies called sub-carriers. Each signal transmitted on a sub-carrier has a different signal strength and phase.
Typical wireless cards provide only a single received signal strength (RSS) value representing the superposition of  information from all sub-carriers. 

Recently, some IEEE 802.11n standard based cards available in the market provide detailed magnitude and phase information about the different sub-carriers represented as \emph{Channel State Information} (CSI). In particular, the  Intel 5300 card reports the CSI for 30 groups of sub-carriers, which is about one group for every 2 sub-carriers for the 20MHz channels operating on the 2.4GHz frequency \cite{Halperin_csitool}.


The IEEE 802.11n nodes also use another technology which is Multiple-Input Multiple-Output (MIMO) where there are multiple transmitter and receiver antennas. Each combination of receiver and transmitter antennas can be considered as a separate virtual link/stream. Therefore, MIMO technology provides multiple virtual streams between a transmitter-receiver pair and hence has the potential of providing better accuracy. 

\subsection{CSI Observations}
\label{sec:csi_prop}
In this section, we show some of our observations on the CSI that we base the system components on. For space constraints, we focus on the magnitude of the received signal strength (RSS) in this paper and leave the phase information to a future paper. 


\begin{figure}[!t]
\centering
        \begin{subfigure}[t]{0.23\textwidth}
                \centering
                \includegraphics[width=\textwidth]{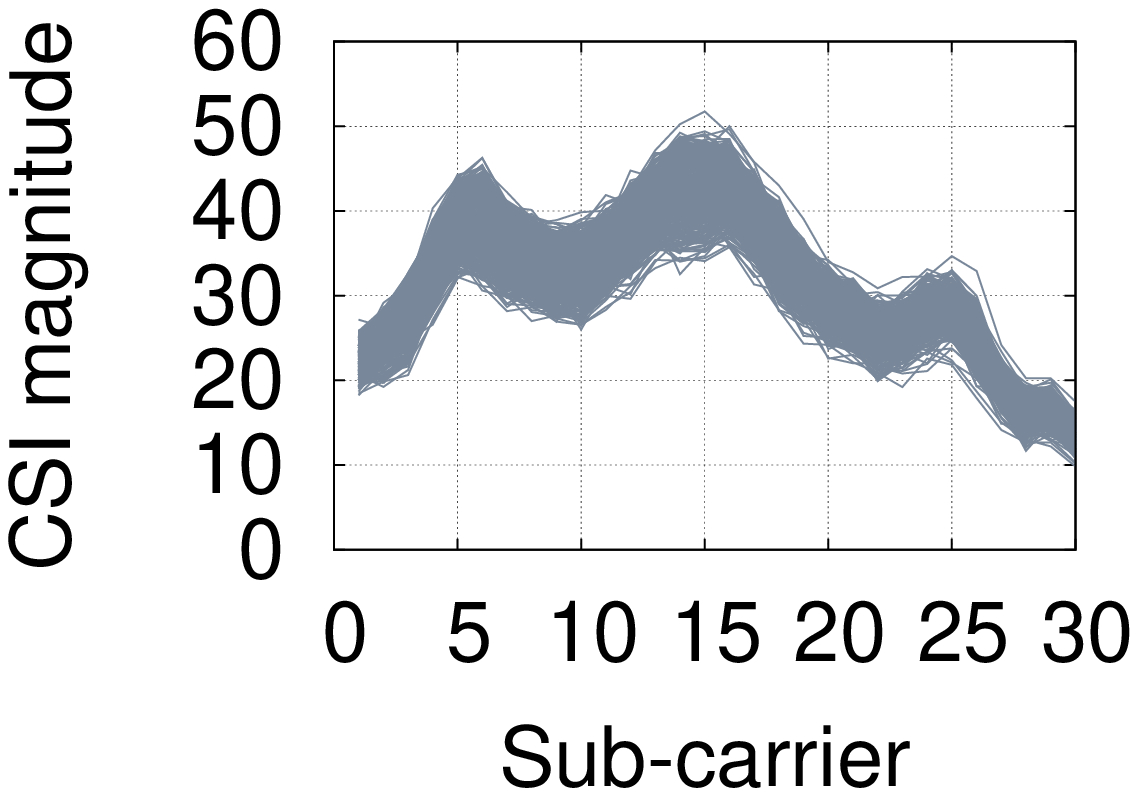}
                \caption{Link with one cluster.}
                \label{fig:one_c}
        \end{subfigure}
        \begin{subfigure}[t]{0.23\textwidth}
                \centering
                \includegraphics[width=\textwidth]{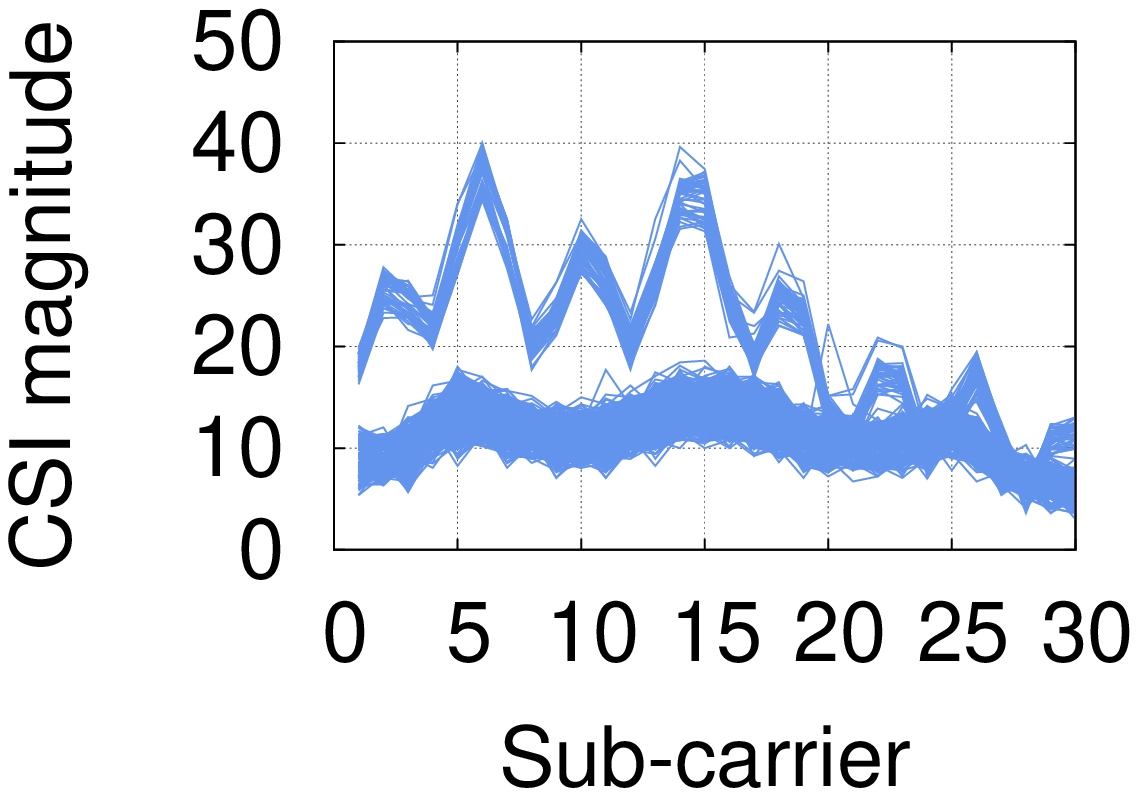}
		\caption{Link with two clusters.}
                \label{fig:two_c}
        \end{subfigure}

	\caption{CSI profile for different virtual links (a link is one transmitter-receiver antennas pair). Each line represents the CSI magnitude of one packet over all sub-carriers. Different lines represent different packets.}
       \label{fig:clustres}
\end{figure}

\begin{figure*}[!t]
\centering
	\begin{subfigure}[b]{0.3\textwidth}
                \centering
                \includegraphics[width=\textwidth]{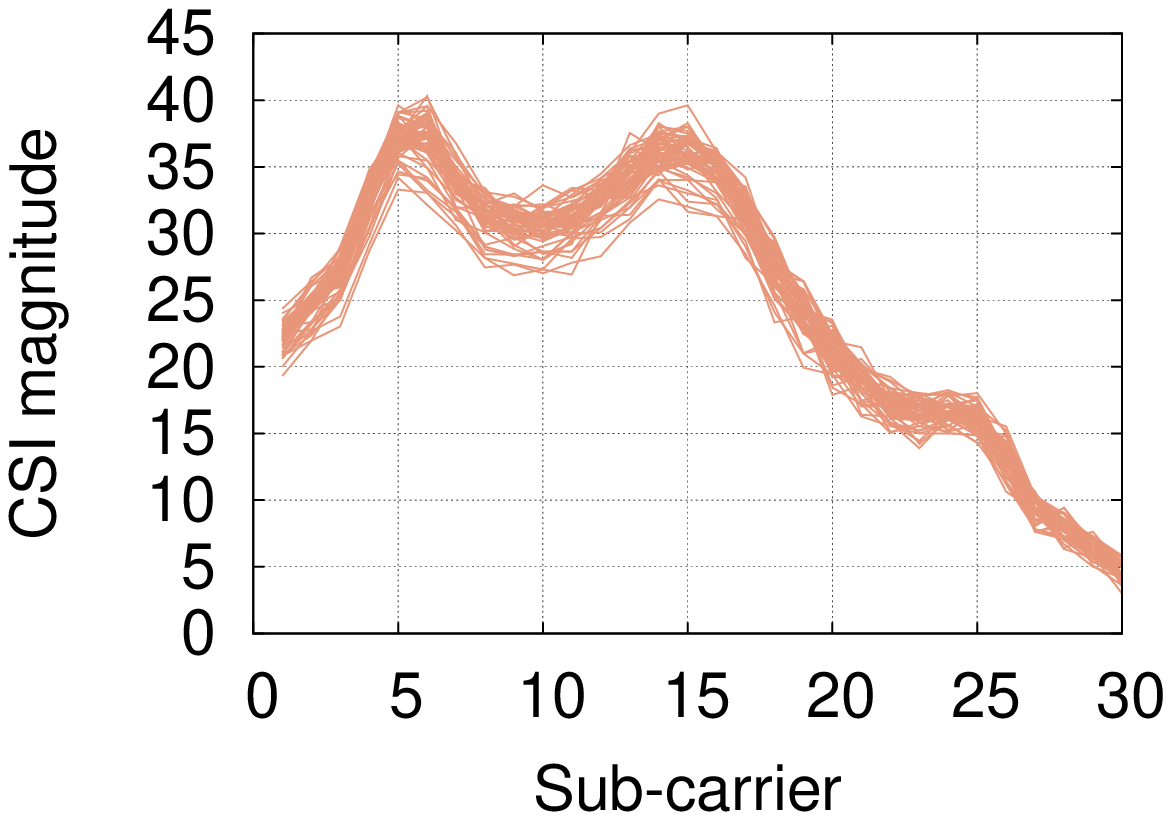}
                \caption{Silence (no human)}
        \end{subfigure}
        \begin{subfigure}[b]{0.3\textwidth}
                \centering
                \includegraphics[width=\textwidth]{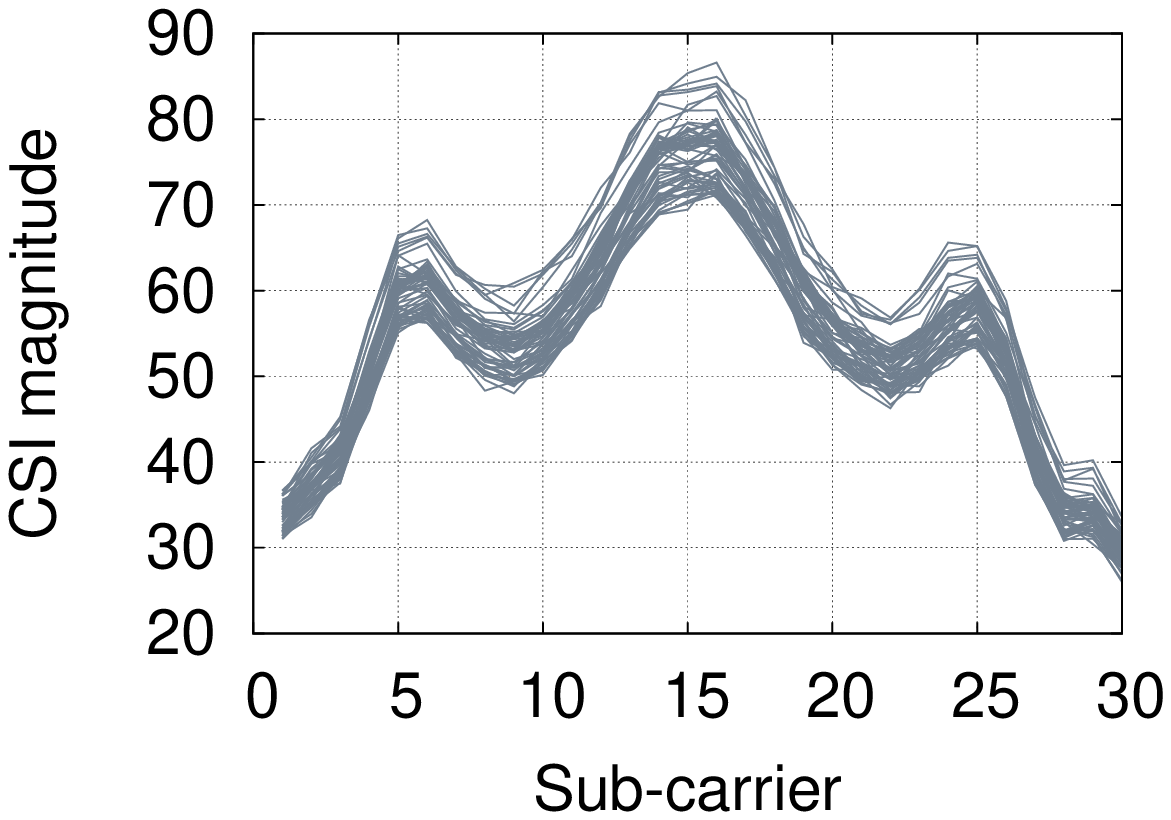}
		\caption{Human at Location 1}
        \end{subfigure}
        \begin{subfigure}[b]{0.3\textwidth}
                \centering
                \includegraphics[width=\textwidth]{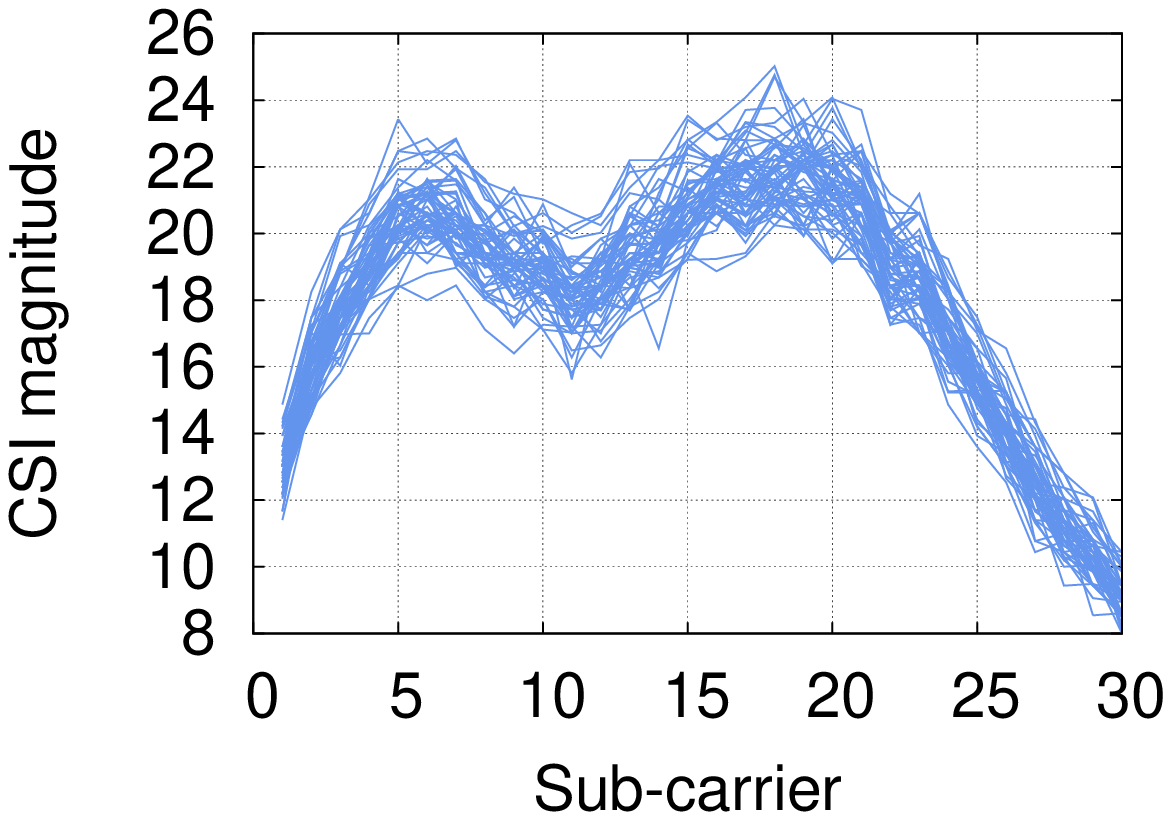}
                \caption{Human at Location 2}
        \end{subfigure}
	\caption{CSI magnitudes for different cases.}
       \label{fig:human}
\end{figure*}

\begin{figure}[!t]
\centering
  \includegraphics[width=0.35\textwidth]{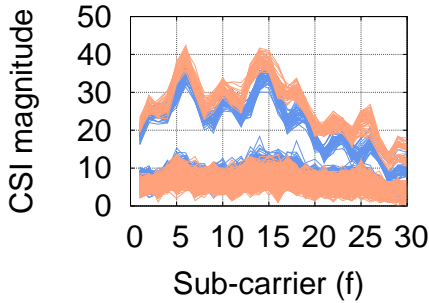}
%

	\caption{CSI information for two adjacent locations (separated by $1m$) overlapped on the same figure.}
       \label{fig:human_adjacent}
\end{figure}



Figure~\ref{fig:clustres} shows the CSI magnitude (profile) for one virtual link, i.e. \emph{a transmitter antenna-receiver antenna pair}, over different packets (each packet is represented by a line) for the 30 sub-carriers. We can notice from the figure that the CSI values for each stream form one or more clusters.

Figure~\ref{fig:human} shows the CSI profile for the silence case as well as the presence of the human at two separated locations for one virtual link. The figure shows that the CSI magnitude information can be used to identify the human presence as well as determine her location.

Figure~\ref{fig:human_adjacent} shows the CSI profile for two adjacent locations (separated by 1 meter) for one virtual link. The figure shows that there is still some difference in the CSI magnitude information between the two locations.  Although the RSS distribution of each sub-carrier of each cluster at each location can be modelled by a Gaussian mixture \cite{Tse:Wireless05,Heba:monophy13}, the \emph{large overlap between the CSI profiles and the smoothness effect}  introduced by the Gaussian mixture leads to \emph{aliasing} between adjacent locations, which reduces accuracy. Instead, \textit{MonoStream} takes a novel approach of treating the CSI profile of a given link at a given location (e.g. figures \ref{fig:one_c} or \ref{fig:two_c}) as an image and employs object recognition techniques with proven accuracy and efficiency to capture the small variations in the profile between adjacent locations.
%
%

%
%
\section{The MonoStream System}
\label{sec:mono_stream}

In this section, we give the details of \textit{MonoStream}. We start by an overview of the system architecture followed by the system model, features construction, and system details.

\subsection{Overview}


\begin{figure*}[!t]
\centering
\includegraphics[width=5in]{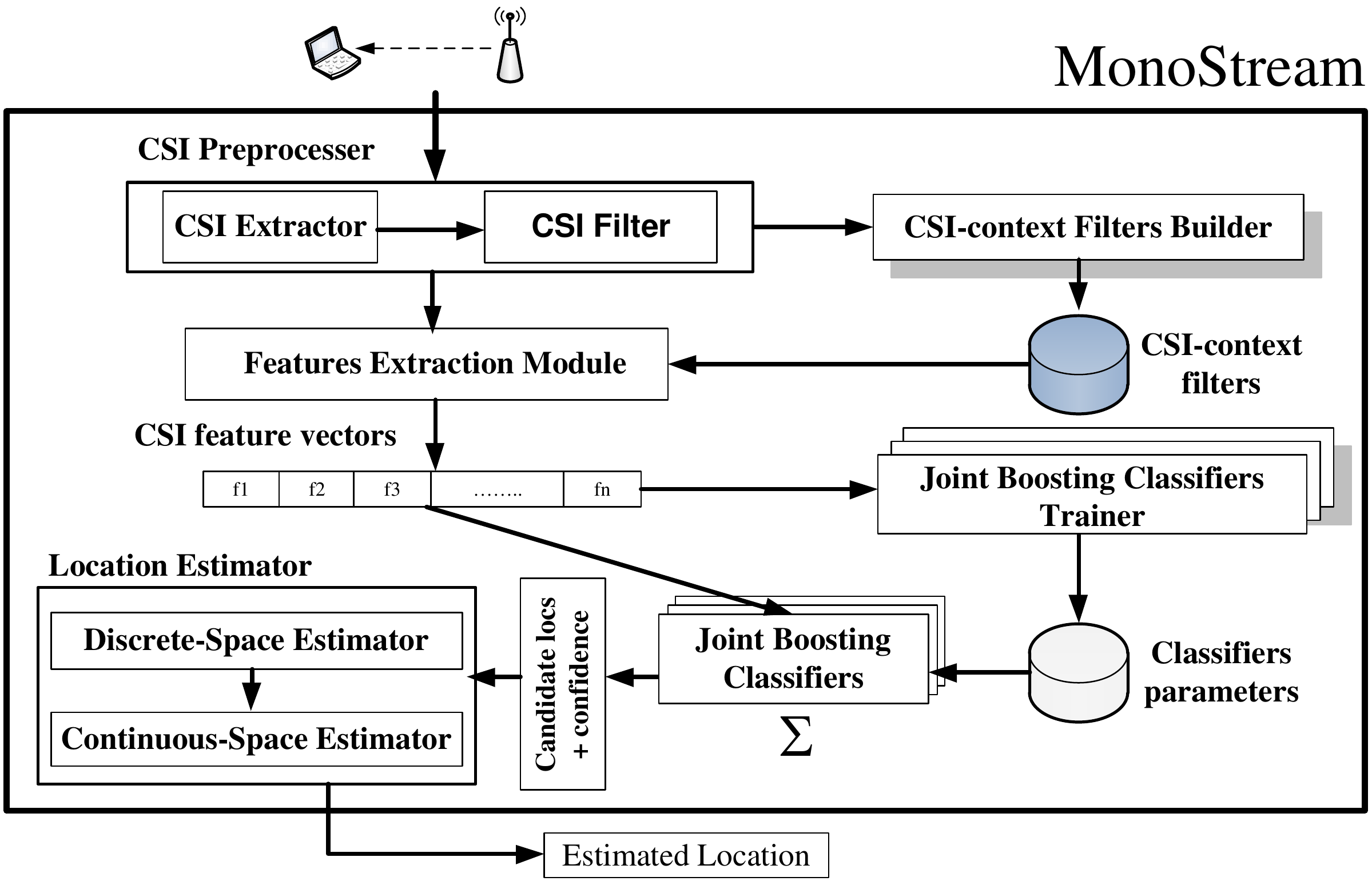}
\caption{\emph{MonoStream} system architecture. Shaded blocks represent modules that run in the offline phase.}
\label{fig:architecture_1}
\end{figure*}

Figure~\ref{fig:architecture_1} shows the system architecture. We have two phases of operation: offline and online. During the \textbf{offline phase}, a person stands at different locations in the area of interest. For each location, CSI values are recorded for all transmitter-receiver pairs (i.e. virtual links) and used to extract features and to train a set of classifiers.
During the \textbf{online phase}, the system uses the collected CSI information to estimate the persons' unknown location. The system is implemented through a set of modules:

The \textbf{CSI Preprocesser} extracts CSI magnitudes from sent packets for each virtual link and filters outlier values.

The \textbf{CSI-context Filters Builder} constructs the filters (during the offline phase) used by the \textbf{Features Extraction Module} to extract the features from the CSI information. These features are used by the \textbf{Joint Booster Classifier Trainer} during the offline phase to efficiently train a set of binary classifiers, one for each location in the area of interest.

The trained classifiers are used during the online phase to provide a list of active locations with positive human detection along with the associated confidence. 

Finally, the \textbf{Location Estimator} fuses the output of the different classifiers to estimate the entity location both in the discrete and continuous space.\\






\subsection{System Model}

\begin{figure}[!t]
\centering
\includegraphics[width=3.2in]{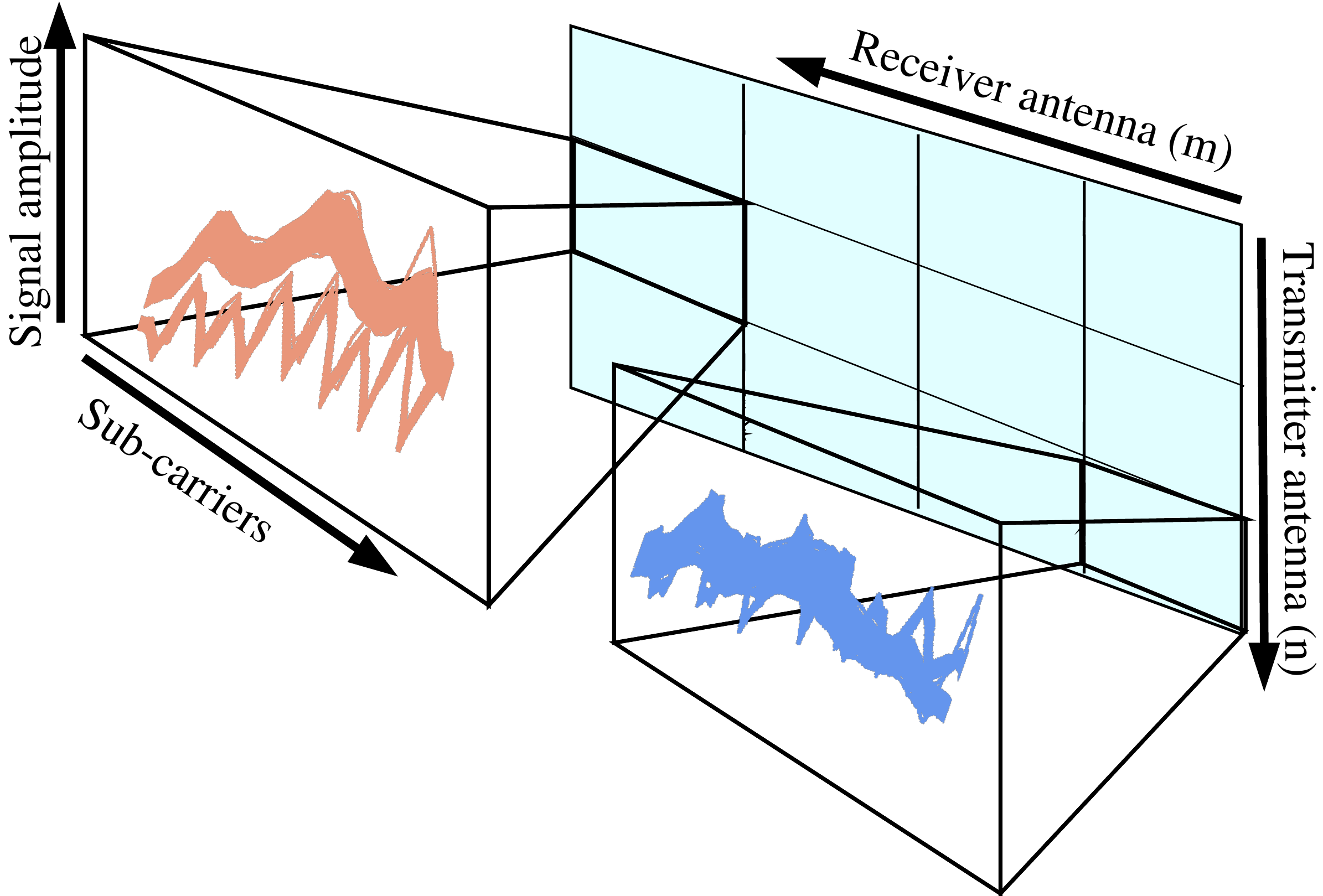}
\caption{Matrix representing CSI information (fingerprint data) at a particular location from all virtual links over all carriers. The matrix indices represent the transmitter and receiver antennas while the content represents the CSI profile.}
\label{fig:matrix}
\end{figure}

Assume an area of interest divided into $L$ fingerprint, i.e. training, locations. This area is covered by only one AP with $n$ MIMO antennas and one MP (e.g. a laptop) with a wireless card having $m$ MIMO antennas. This leads to $n.m$ \emph{virtual links} between the transmitter and receiver. Using the OFDM modulation, each transmitted packet is sent using $f$ sub-carriers on each of the $n$ antennas. \footnote{This leads to a total of $n.m.f$ \emph{virtual streams} between each physical transmitter-receiver pair, as compared to only only stream with the traditional RSS-based techniques \cite{Nuzzer_TMC,Seifeldin:Nuzzer}.}


During the training phase, a human stands at each location $l \in L$ and the CSI of a set of packets sent by the AP are recorded at the MP. The CSI information of all packets corresponding to location $l$ is represented by a matrix (Figure~\ref{fig:matrix}) of size $n.m$, whose entry $(i,j)$ represents the CSI profile (as in Figure~\ref{fig:clustres}) of the virtual link representing the packets from transmitter antenna $i$ to receiver antenna $j$. This rich information is what allows \emph{MonsStream} to achieve high accuracy with just a single AP-MP pair.


%

\subsection{Features and Context Filters}
\label{sec:spotons_context_filters}

\begin{figure}[!t]
\centering
\includegraphics[width=3.2in]{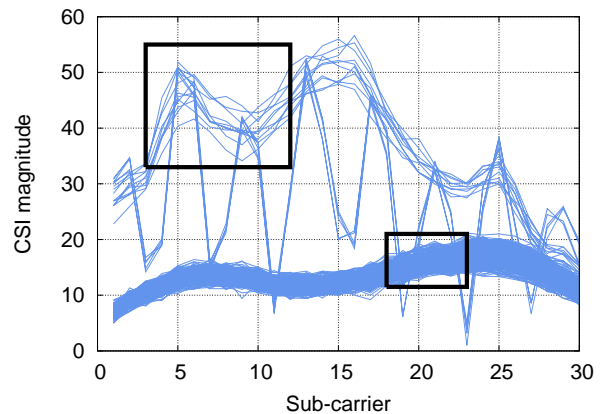}
\caption{Context-filter examples: each filter is represented by a rectangle of random size over the CSI profile of a single virtual link (transmitter-receiver pair).}
\label{fig:context_filter}
\end{figure}

This section discusses our proposed \textit{CSI-based} features that will be used to train the discrimination model used to differentiate between different locations. To capture small variations in the CSI profiles between adjacent locations, \emph{MonStream} borrows techniques from the object recognition domain \cite{Leung:Texton01,Shotton:TextonBoost06}, where the CSI profile of a particular virtual link at a certain location is treated as an image. Random sampling of features within these
CSI-profile images is used to reduce complexity as well as increase the discrimination between adjacent locations.

In particular, we define a \emph{context filter} on a CSI profile as a rectangular mask whose width is determined by the range of sub-carriers covered; and height determined by the range of CSI magnitude values covered (Figure~\ref{fig:context_filter}). For example, we can define a context filter that covers sub-carriers from 3 to 12, and CSI magnitudes from 32 to 55. A set of $d$ context filters $T = \{t_i\}; i = 1, ... , d$ is then defined by selecting rectangles with different random sizes (width and hight uniformly distributed over the range) over the CSI profiles in the matrix. These random filters, once selected, are then fixed over the different locations. 


Let $\lambda_t$ be the number of packets that fall inside a \textit{CSI-context} filter $t \in T$. Instead of using  $\lambda_t$'s directly, \emph{MonoStream} uses Haar-like \cite{Paul:Haar01} features\footnote{The key advantage of a Haar-like feature over other features is its calculation speed.} to increase the feature space and hence lead to better accuracy. Specifically, for \emph{two} different CSI-context filters $t_i, t_j \in T$, the associated feature $\lambda_{i,j}$ is obtained as $\lambda_i- \lambda_j$. Note that there are $d \choose 2$ different combinations of features to choose from, each representing a pair of filter indices $(i, j)$. The choice of $d$ represents a trade-off between accuracy and efficiency as quantified in the Evaluation Section. This feature space is further reduced by selecting the best features during the classifier training phase.


\subsection{Classifier Training Module}
\label{sec:class_train}
Once the features are extracted, the goal of the \emph{Classifier Training Module} is to train a binary classifier for each location to differentiate it from all other fingerprint locations based on the training data. To balance computational efficiency and accuracy, we adopt an AdaBoost-based classifier \cite{Freund:1995,Mahamud:Boosted01,Torralba:Sharing04}. AdaBoost is an iterative technique that adds a new simple weak classifier in each iteration. At each round, samples that are misclassified from previous rounds are assigned higher weights so that the current weak classifier focuses on disambiguating them. A weight is assigned to each
classifier and the final classifier is defined as the linear combination of the classifiers from each
iteration. AdaBoost can be regarded as a feature selection technique that selects the most discriminant $g$ features, where $g$ is the number of iterations. Therefore, this further reduces the number of features, and hence is more computationally efficient.

We adopt decision stumps as our weak binary classifiers. A decision stump is defined by three parameters: a feature that it classifies, a threshold on the feature value, and the sign of the decision (whether it takes a positive or negative value if the feature value is above the threshold). Therefore, implementing our weak classifier requires just one condition check, which is extremely efficient.

Furthermore, instead of independently training the classifier of each location, we use a joint boosting approach that selects the best features that can be used to discriminate between the largest subset of locations \cite{Torralba:Sharing04} in each iteration. For a given performance level, joint boosting reduces the training overhead to logarithmic in the number of locations (as features are shared between locations), increases accuracy, reduces running time, and avoids over-fitting of the training samples. This last property is of specific importance in DF localization due to the noisy and dynamic nature of the wireless channel.
%


\subsection{Location Estimator}
The purpose of this module is to estimate the actual user location given a received CSI vector/profile $S$. We start by assuming that the user is standing at one of the discrete fingerprint locations then we generalize this to an arbitrary location in the next subsection.

\subsubsection{Discrete-space estimator}
The extracted features from the CSI vector are processed by the \emph{Joint Boosting Classifiers} to generate the binary decision for each location and its associated confidence. For a classifier for fingerprint location $l \in L$, let $d_l$ be 1 if the classifier detects that a person is present at this location and -1 otherwise. In addition, let $c_l$ the decision confidence of that classifier. In order to fuse the output of the different classifiers, we adopt a probabilistic approach where we want to find the location $l^*$ in the fingerprint $L$ that maximizes the probability $P(l|S)$. That is:
\begin{equation}
  l^*= \arg\max_l P(l|S)
\end{equation}

Using Baysian inversion, this can be represented as:

\begin{equation}
  l^*= \arg\max_l \frac{P(S|l).P(l)}{P(S)}
  \label{eq:inversion}
\end{equation}

Assuming all locations are equally likely\footnote{If the probabilities distribution of $P(l)$ is known, it can be used directly in Equation~\ref{eq:inversion}} and noting that $P(S)$ is independent of $l$, Equation~\ref{eq:inversion} becomes:

\begin{equation}
  l^*= \arg\max_l P(S|l)
  \label{eq:final_inv}
\end{equation}

We estimate $P(S|l)$ 
as a function of the classifiers confidence for the location with positive detection as:
\begin{equation}
P(S | l) = \frac{c_l}{\sum\limits_{i; \forall d_i=1} c_i}
\label{eq:soft_max}
\end{equation}

To further enhance the accuracy, \emph{MonoStream} calculates the probability based on a sequence of packets during a time window $w$.

\subsection{Continuous-space estimator} \label{sec:cont}
The previous estimator will always return one of the fingerprint locations, even if the entity is standing in a location that does not coincide with any of the fingerprint locations. To further enhance accuracy, the continuous space estimator uses a spatial averaging technique.
The spatial averaging technique estimates the location as the weighted average of the most probable $k$ locations, where each location is weighed by its probability (Equation~\ref{eq:soft_max}) normalized by the sum over all probabilities.
Note that using the continuous-space estimator, \textit{MonoStream} can \textbf{\emph{achieve accuracy that is better than the fingerprint grid spacing}}.
%

\section{Performance Evaluation}
\label{sec:eval}

\begin{figure}[!t]
\centering
      \begin{subfigure}[b]{0.32\textwidth}
                \centering
                \includegraphics[width=\textwidth]{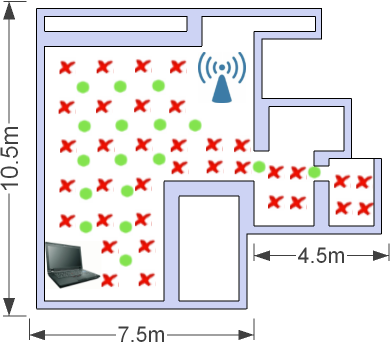}
                \caption{Testbed 1.}
                \label{fig:map}
     \end{subfigure}

	\begin{subfigure}[b]{0.25\textwidth}
                \centering
                \includegraphics[width=\textwidth]{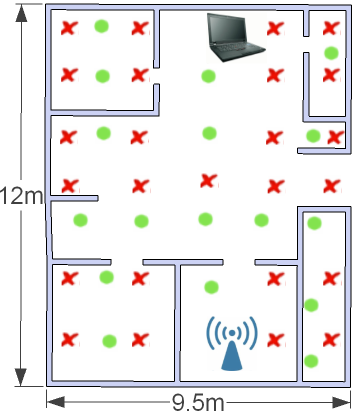}
                \caption{Testbed 2.}
                \label{fig:layout2}
     \end{subfigure}           
     \caption{Testbeds for the experiment with training locations marked as red crosses and testing locations marked as green circles.}
     \label{fig:testbeds}
\end{figure}

In this section, we analyze the performance of \emph{MonoStream} and compare it to the state-of-the-art DF WLAN localization systems \cite{Seifeldin:Nuzzer,Nuzzer_TMC, Heba:monophy13}. We start by describing the experimental setup and data collection. Then, we analyze the effect of different parameters on the system performance. We end the section by a comparison with the state-of-the-art.

\subsection{Testbeds and Data Collection}

We conducted two experiments with different testbeds (Figure~\ref{fig:testbeds}). The first testbed covers a typical apartment with an area of approximately $100 m^2$ (about 1077 sq. ft.) while the second testbed represents a residential apartment with an area of $114 m^2$ (about 1228 sq. ft.). Both testbeds are with typical furniture. The area was covered by a single Cisco Linksys X2000 AP and a Dell Adamo XPS laptop as a MP. The laptop has an Intel 5300 card that can provide CSI information \cite{Halperin_csitool}. The fingerprint is constructed for 35 (25) different locations for the first (second) testbed, uniformly distributed over the testbed area. An independent test set of 17 locations are chosen randomly between the training locations at different times of day using different persons from the training set. Table~\ref{tab:def_val} shows the default values for the different parameters.  For the \textbf{running time} estimation, all experiments were performed on a Dell Latitude E6510 with an Intel Core~i7 CPU running at  2.67 GHz and having 8GB RAM. Due to space constraints, we give the details of the first testbed and summarize the results of the second.\\

\begin{table}[!t]
    \centering
\begin{tabular}{|c|p{1cm}|l|} \hline
      Parameter & Default value & Meaning\\ \hline \hline
         $m$ & 2  & Num. of receiver antennas\\ \hline
        $f$ & 30  & Num. of  sub-carriers\\ \hline
		$g$ & 700  & Num. of  boosting rounds\\	\hline
         $w$ & 500  & Num. of packets per loc. est.\\ \hline
        $k$ & 6  & Spatial avg. window\\ \hline
  \end{tabular}
    \caption{Default parameters values.}
    \label{tab:def_val}
\end{table}

\subsection{Effect of Different Parameters}

\subsubsection{Effect of the number of receiver antennas ($m$)}
\begin{figure}[!t]
\centering
\includegraphics[width=0.5\textwidth]{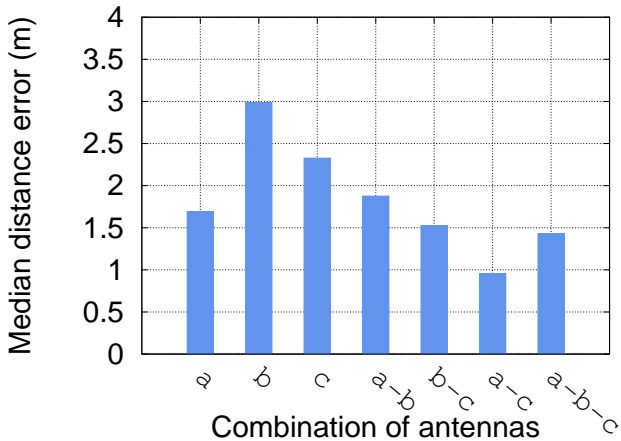}
\caption{Effect of different combinations of receiver antennas $(a, b, c)$.}
\label{fig:cdf_ant}
\end{figure}

Figure \ref{fig:cdf_ant} shows the effect of changing the antennas combinations on the median distance error. The figure shows that different combinations lead to different accuracy. This is due to the noisy wireless channel and the different multipath effects encountered by the packets received at the different antennas. This means that using more antennas does not necessarily lead to better accuracy. The good news is that the SNR associated with the antennas can be used to determine the best combination. For the rest of this section, we use antennas a and c (i.e. $m=2$) as they lead to the best accuracy.

\subsubsection{Effect of number of packets used in estimation ($w$)}

\begin{figure}[!t]
\centering
\includegraphics[width=0.5\textwidth]{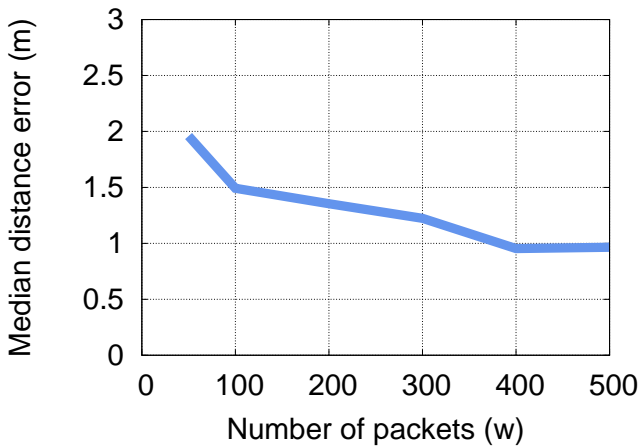}
\caption{Effect of the number of packets used in one location estimate ($w$).}
\label{fig:cdf_packets}
\end{figure}

 Figure \ref{fig:cdf_packets} shows the effect of increasing the number of packets used in the estimation process. The figure shows that as the number of packets increases, the accuracy increases. However, increasing $w$ increases the latency.  Therefore, there is a tradeoff that a designer needs to balance based on her needs. Setting $w = 500$ gives high accuracy of $0.95m$ with reasonable latency.

\subsubsection{Effect of processed sub-carriers ($f$)}

\begin{figure}[!t]
\centering
\includegraphics[width=0.5\textwidth]{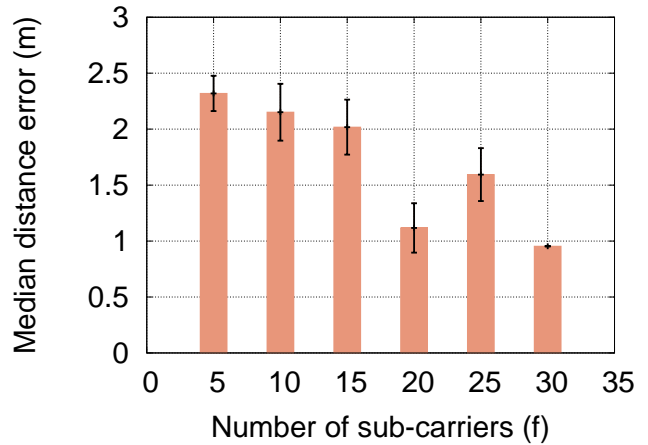}
\caption{Effect of the number of sub-carriers used. Error bars represent the standard deviation over different randomly selected sets of sub-carriers.}
\label{fig:cdf_subc}
\end{figure}

Figure \ref{fig:cdf_subc} shows the effect of increasing the number of sub-carriers on the median distance error. For a specific number of sub-carriers ($f$), we choose a number of random subsets of size $f$ from the available 30 sub-carriers and draw both the average and standard deviation of performance. The figure shows that, in general, increasing the number of sub-carriers leads to better accuracy. This due to the increased amount of available information. Therefore, a designer can tune the the accuracy-computational complexity if needed.

\subsubsection{Effect of number of averaged locations ($k$)}

\begin{figure}[!t]
\centering
\includegraphics[width=0.5\textwidth]{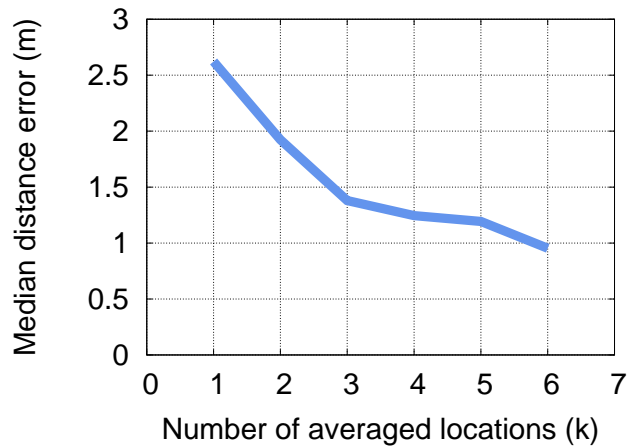}
\caption{Effect of the number of averaged locations ($k$).}
\label{fig:k_avg}
\end{figure}

Figure \ref{fig:k_avg} shows the effect of increasing the number of averaged locations ($k$) for the continuous-space estimator. The figure shows that increasing the number of averaged locations reduces the median distance error until it saturates around $k=3$.

\subsubsection{Effect of number of CSI-context filters ($d$)}

\begin{figure}[!t]
\centering
\includegraphics[width=0.5\textwidth]{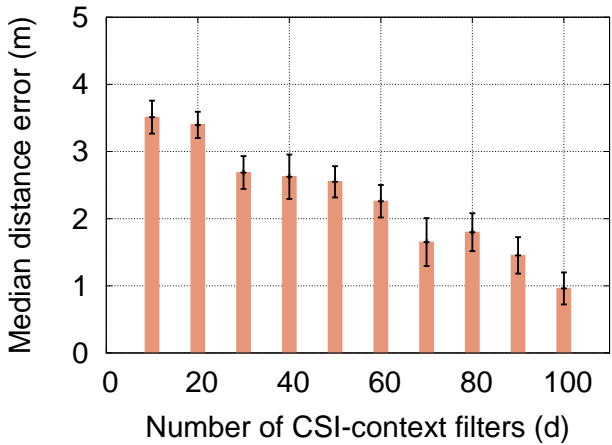}
\caption{Effect of the number of CSI-context filters ($d$). Error bars represent the standard deviation over 10 different random selection of filters.}
\label{fig:features_num_effect}
\end{figure}

Figure \ref{fig:features_num_effect} shows the effect of increasing the number CSI-context filters. The figure shows that increasing the number of CSI-context filters reduces the median distance error. We pick $d = 100$ as it balances accuracy and computational overhead.

\subsubsection{Effect of number of boosting rounds ($g$)}
    
\begin{figure}[!t]
     \centering
	\begin{subfigure}[b]{0.5\textwidth}
                \centering
    \includegraphics[width=\textwidth]{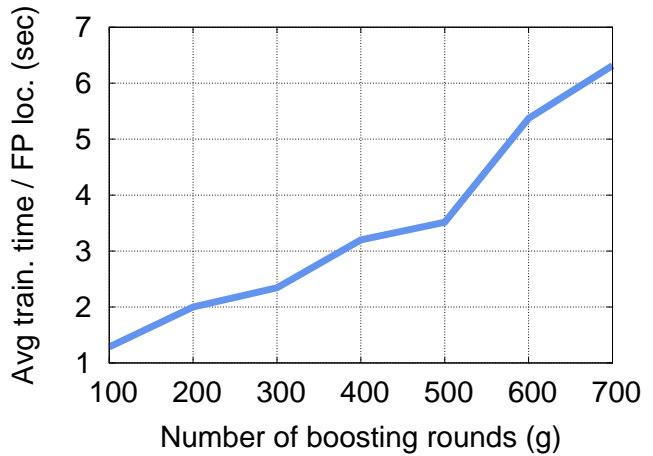}
    \caption{Offline training phase.}
    \label{fig:boosting_offline}
    \end{subfigure}

    \begin{subfigure}[b]{0.5\textwidth}
    \centering
    \includegraphics[width=\textwidth]{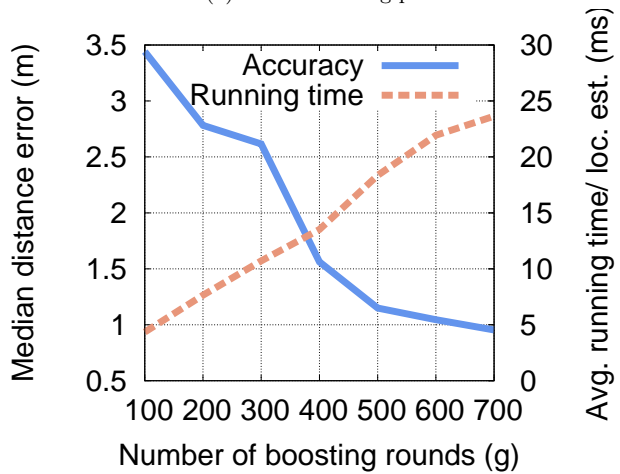}
    \caption{Online phase.}
    \label{fig:boosting_online}
    \end{subfigure}
    \caption{Effect of the number of boosting rounds on performance.}
    \label{fig:boosting}
\end{figure}

Figures~\ref{fig:boosting_offline} and \ref{fig:boosting_online} show the effect of the number of boosting rounds on the training time and online phase performance respectively. The figure shows that the training and estimation time increases linearly with the boosting rounds. Simultaneously, the accuracy increases with the increase of the boosting rounds until it saturates at $R=700$. Therefore, we take this value as the default value.

 
\subsection{Comparison with the State-of-the-Art}

\begin{figure}[!t]
\centering
\centering
        \begin{subfigure}[b]{0.5\textwidth}
                \centering
                \includegraphics[width=\textwidth]{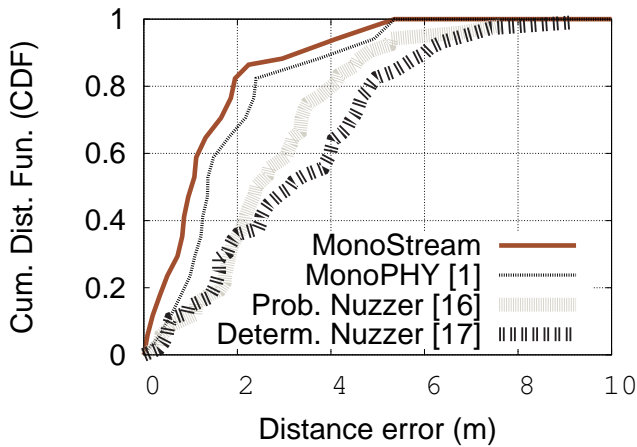}
                \caption{CDF of distance error.}
                \label{fig:cdf_streams}
        \end{subfigure}

        \begin{subfigure}[b]{0.5\textwidth}
                \centering
                \includegraphics[width=\textwidth]{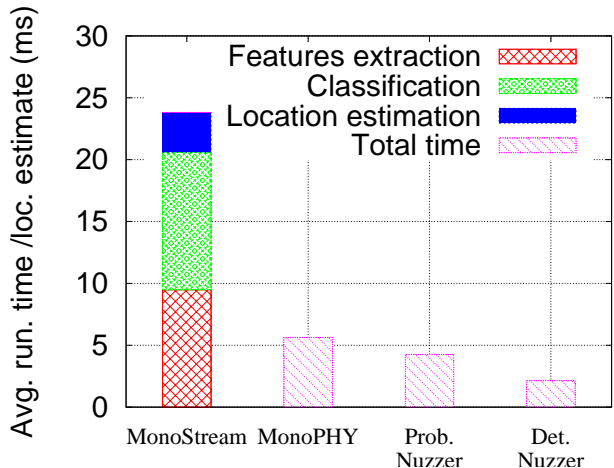}
                \caption{Running time.}
                \label{fig:run_time}
        \end{subfigure}
\caption{Comparison with the state-of-the-art techniques.}
\label{fig:comp}
\end{figure}

Figure~\ref{fig:cdf_streams} compares the CDF of distance error for the \emph{MonoStream} system to the Deterministic \cite{Seifeldin:Nuzzer} and Probabilistic \emph{Nuzzer} \cite{Nuzzer_TMC} traditional DF systems\footnote{These techniques do not use the CSI information but rather the combined RSS only as reported by the MAC layer.} designed for multiple streams as well as the \emph{MonoPhy} system~\cite{Heba:monophy13} designed for a single stream based on a Gaussian mixtures approach. Table~\ref{tab:results_summary} summarizes the results. The results show that \emph{MonoStream} has the best accuracy with an enhancement of at least $26.3\%$ in median distance error over the best state-of-the-art technique \textbf{in a single stream environment}.

Figure~\ref{fig:run_time} shows the running time of the different technique. The figure shows that the accuracy advantage of \emph{MonoStream} does not compromise efficiency; Only $23ms$ are required per location update, which shows that  \emph{MonoStream} can run in realtime.\\ \\

\begin{table*}[!t]
    \centering
    \caption{Performance summary for the different systems under the two testbeds using median distance error and running time as the metric. Numbers between parenthesis represent the percentage difference from \emph{MonoStream}.}
\small
 \begin{tabular}{|l||p{1.2cm}|p{1.5cm}||p{1.2cm}|p{1.5cm}|} \hline
       & \multicolumn{2}{c||}{Testbed 1}& \multicolumn{2}{c|}{Testbed 2} \\\cline{2-5}
          & Median &Running &   Median &Running \\
       \raisebox{3ex}{System} & \raisebox{1.5ex}{error} & \raisebox{1.5ex}{ time}&\raisebox{2ex}{error}  &\raisebox{1.5ex}{time}\\ \hline \hline
         MonoStream &   0.95m &  23.64ms & 1.4m & 16.28ms\\ \cline{1-5}
         MonoPHY \cite{Heba:monophy13}&  1.36m (30\%) &  5.62ms (76.22\%) & 1.9m (26.3\%)& 4.44ms (72.73\%)\\ \hline \hline
       Prob. Nuzzer \cite{Nuzzer_TMC}& 2.63m (63.87\%)& 4.26ms (81.98\%)& 3.1m (54.83\%) &  3.47ms (78.68\%)\\ \hline
       Det. Nuzzer \cite{Seifeldin:Nuzzer}&  3.16m (69.93\%)& 2.14ms (90.94\%) & 3.9m (64.1\%)&  1.67ms (89.74\%)\\ \hline    \end{tabular}
\label{tab:results_summary}
 \end{table*}

\section{Related Work}
\label{sec:related}

Device-free tracking schemes have advanced over the year including: radar-based, camera-based, sensors-based, and WLAN-based systems.
In the radar-based systems, pulses of radio waves are transmitted into the area of interest and based on measuring the received reflections, objects could be tracked. Several approaches have been presented in this class including ultra-wideband (UWB) systems \cite{Yang:UWB}, doppler radar \cite{Lin:Doppler}, and MIMO radar systems \cite{Haimovich:MIMO}.
On the other hand, camera-based tracking systems are based on analyzing a set of captured images to estimate the current locations of objects of interest \cite{Moeslund:Camera,Krumm:Camera}. Sensor-based systems use especially installed sensor nodes to cover the area of interest. For example, \cite{Patwari:RTI} applies radio tomographic techniques to the readings of a dense array of sensors to obtain accurate DF tracking.

All these technologies share the requirement of installing special hardware to be able to perform DF tracking, which reduces their scalability in terms of cost and coverage area. In contrast, WLAN DF tracking aims at exploiting the already installed WLAN.
DF localization in WLANs was first introduced in \cite{Youssef:DFPchallenges} along with feasibility experiments in a controlled environment. Several papers followed the initial vision to provide different techniques for detection and tracking of a single entity \cite{Moussa:smart,Yang:Performin,Kosba:RASID,Seifeldin:Nuzzer}.
 Tracking multiple entities was proposed in \emph{SPOT} \cite{Sabek:Spot12} based on a probabilistic energy minimization framework that combines a conditional random field with a Markov model to capture both of spatial and temporal relations between active locations. All these techniques rely only on the reported signal strength from the MAC layer and hence require a large number of streams, i.e. APs and MPs, to provide acceptable accuracy.

The closest work to ours is the \emph{MonoPhy} system \cite{Heba:monophy13} that models the CSI information using a Gaussian mixtures approach. However, since the CSI profiles in adjacent locations are usually similar, the smoothness effect of the Gaussian mixture leads to aliasing in the CSI space between adjacent locations, which reduces accuracy.

The novel approach adopted from the object recognition domain allows the \emph{MonoStream} system to capture the small variations in RSS between adjacent locations and maintain high efficiency.

\section{Discussion}
\label{sec:discussion}

\subsection{Accuracy}

\emph{MonoStream} accuracy is based on three factors: (1) leveraging the rich CSI information, (2) sampling a large number of features that capture small variations between adjacent locations, and (3) using a joint boosting technique that selects the combined best features over all locations and avoids over training. These factors allow it to tolerate a low number of streams and achieve high accuracy.

\subsection{Efficiency}
The increase in accuracy of \emph{MonoStream} does not come at an increased complexity. \emph{MonoStream} employs a number of techniques to enhance its efficiency and trade the accuracy with efficiency as quantified in the evaluation section. These include: (1) using easy-to-compute Haar-like features, (2) using decision stumps that can be computed using only a single comparison check, and (3) adopting an joint boosting approach that selects the best features, reduces training overhead and reduces the feature space. This in turn reflects in reducing the running time of the algorithm. 
Theoretically, the training complexity of the joint boosting classifier is $O( g.|L|^2)$ \cite{Torralba:Sharing04} whereas the location estimation complexity is $O(g.|L|)$.

\subsection{Multi-entity Tracking}
Although all results in the paper focus on a single entity, the extension to the multiple entities case is straight forward. Current approaches for multi-entity DF localization, e.g. \cite{Sabek:Spot12}, can be applied directly to the \emph{MonoStream} system.

\subsection{Dynamic Changes in the Environment}
Another important aspect of the practical deployment of \emph{MonoStream} is handling the dynamic changes in the environment which may require re-calibration of the area of interest. Different approaches can be applied to capture these dynamic changes including dynamically updating the stored parameters, e.g. using anomaly detection techniques as in \cite{Kosba:RASID}, and using CAD tools for DF systems, e.g. as in \cite{AROMA}.

\section{Conclusion}
\label{sec:conclusion}
We tackled the device-free passive localization problem using physical layer information supported by WLAN standards. We presented the \emph{MonoStream} system based on a novel set of CSI-context features that can capture minimal variations in the CSI profiles between adjacent locations. Combined with an efficient AdaBoosting classifier, this allows us to achieve both accurate and efficient DF localization using only a single transmitter and receiver. 

Experimental evaluation in two typical WiFi testbeds shows that \emph{MonoStream} can achieve $0.95 m$ median distance error, which is better than the state-of-the-art techniques by at least 26\%. This enhancement in accuracy comes with an efficient execution of less than $23ms$ per location update on a typical laptop. This highlights the promise of \emph{MonoStream} for real-time DF tracking applications.

Currently, we are expanding \emph{MonoStream} in multiple directions including integrating the CSI phase information, multiple entities tracking, and entity identification.


\normalsize
\bibliographystyle{abbrv}
\bibliography{DFPRefer}

\balance

\end{document}